\documentclass[12pt,showkeys,amsmath,amssymb]{revtex4}
\usepackage{amsmath,amsfonts,amsthm,amscd,amssymb,latexsym}
\usepackage{bm}
\usepackage{dcolumn}
\usepackage{graphicx}
\usepackage{epstopdf}
\usepackage{color}
\usepackage{epsf}
\usepackage{epsfig}
\usepackage{graphicx, epic, eepic, color}
\usepackage{soul}
\usepackage{tikz}
\usetikzlibrary{calc}
\usetikzlibrary{decorations.markings}
\usetikzlibrary{shadings, calc, decorations.markings}

\def\beq{\begin{equation}}
\def\eeq{\end{equation}}
\begin{document}
\title{Dynamical Friction and Tidal Interactions}

\author{Mahmood \surname{Roshan}$^{1,2}$}
\email{mroshan@um.ac.ir}
\author{Bahram \surname{Mashhoon}$^{2,3}$}
\email{mashhoonb@missouri.edu}

\affiliation{$^1$Department of Physics, Faculty of Science, Ferdowsi University of Mashhad, P.O. Box 1436, Mashhad, Iran \\
$^2$School of Astronomy, Institute for Research in Fundamental Sciences (IPM), P. O. Box 19395-5531, Tehran, Iran\\
$^3$Department of Physics and Astronomy, University of Missouri, Columbia, Missouri 65211, USA\\
}

\begin{abstract}
We discuss dynamical friction in an $N$-body system in the presence of tidal interactions caused by a distant external source. Using the distant tide approximation, we develop a perturbation scheme for the calculation of dynamical friction that takes tidal effects into account to linear order. In this initial analytic approach to the problem, we neglect the influence of tides on the distribution function of stars in the background stellar system. Our result for the dynamical friction force in the appropriate limit is in agreement with Chandrasekhar's formula in the absence of tides.  We provide preliminary estimates for the tidal contributions to the dynamical friction force. The astrophysical implications of our results are briefly discussed. 
\end{abstract}


\maketitle

\section{Introduction}

When a body of mass $M$ moves with velocity $\mathbf{v}_{M}$ through an infinite homogeneous medium of stars with average stellar mass $m$, $m \ll M$, it slows down due to the gravitational drag of the stars. The resulting \emph{dynamical friction force} was first calculated by Chandrasekhar~\cite{Chandrasekhar:1943ys, BT} and is given by
\begin{equation}\label{I1}
M \frac{d\mathbf{v}_M}{dt}  = -4\pi G^2 M^2 \frac{\mathbf{v}_M}{v_\text{M}^3} \rho(< v_M) \ln \Lambda\,,
\end{equation}
where $v_M = |\mathbf{v}_{M}|$, $\rho(< v_M)$ is the density of stars with speeds less than $v_M$ and $\ln \Lambda$ is the so-called Coulomb logarithm. In fact, $\Lambda =  b_{\rm max} / b_{\rm min}$, where $b$ is the scattering impact parameter, $b_{\rm max} = D$  is the diameter of the smallest sphere that completely surrounds the whole system and $b_{\rm min}$ is the maximum of the size of the incident mass $M$ and $GM/v_M^2$. For the background material and a comprehensive treatment of Chandrasekhar's formula~\eqref{I1} and its limitations, see Binney and Tremaine~\cite{BT}. Equation~\eqref{I1} can be interpreted in terms of gravitational wake, namely, the disturbance caused by the motion of $M$ through the medium produces a density enhancement in the incident body's wake. This overdensity decelerates $M$ via gravitational attraction; see~\cite{TW} and the references cited therein.  

Chandrasekhar's formula~\eqref{I1} is a consequence of two-body Newtonian scattering of $M$ from each of the stars in the medium; indeed, it is an approximate result that ignores the attractive gravitational interaction of the stars. In the scattering of $M$ from a star of mass $m$, the final ($t = \infty$) deflected momentum of $M$ has a component along the initial ($t = -\infty$) direction of its incident momentum $M\,\mathbf{v}_{M}$ that is invariant to linear order in the Newtonian constant $G$, but \emph{decreases} to order $G^2$ and beyond in accordance with~\cite{BT}
\begin{equation}\label{I2} 
M\,(\delta \mathbf{v}_M)_{||} =  - 2\,m\,M\,\mathbf{v}_0\,\frac{G^2 (m+M)}{G^2 (m+M)^2 + b^2 v_0^4}\,,  
\end{equation}
where $\mathbf{v}_0 = \mathbf{v} (t = -\infty)$, $\mathbf{v} = \mathbf{v}_M - \mathbf{v}_m$ is the relative velocity, $v_0 = |\mathbf{v}_0|$ and $b$ is, as before, the impact parameter. Here, $M\,(\delta \mathbf{v}_{M})_{||}$ is the net change in the momentum of $M$ along its initial direction of incidence as a consequence of gravitational scattering from $m$. The net loss of momentum of $M$ along its initial direction of motion is the source of the dynamical friction force. 

While numerous studies have verified that Chandrasekhar's formula~\eqref{I1} gives a rather accurate description in many different systems (e.g.,~\cite{bosch,BK}), the derivation of  formula~\eqref{I1}  is  based on the assumption of an infinite homogeneous background medium and furthermore neglects the self-gravity of the background stars. In the meantime, the calculation of dynamical friction has been extended to inhomogeneous media. In spherically symmetric environments, for example, the phenomenon of  ``core stalling" has been found, where dynamical friction disappears within the central constant-density cores of massive halos~\cite{Read:2006fq, bosch2}.  It has also been shown that a body orbiting outside a self-gravitating system experiences orbital decay and loses angular momentum~\cite{lin}. Moreover,
the influence of gravitational interaction of the background stars on dynamical friction has been 
considered in several analytical and numerical studies; see~\cite{TW, bosch2} and the references cited therein. 

In connection with tides, theoretical studies of dynamical friction in the presence of tidal interactions have generally used the impulse approximation for the sake of simplicity and have thus been restricted to the \textit{tidal shocking} method~\cite{BT,spitzer,colpi,gnedin, Banik}. The impulse approximation was originally used by Fermi in quantum scattering calculations~\cite{Fermi}. In reference to tidal interaction, it signifies the change in the velocities of stars as a result of a brief tidal encounter of a galaxy. The time integral of tidal force is the impulse that results in the change of momentum. Within this approximation, the timescale of the encounter between the system and the external tidal source is much shorter than the dynamical timescale associated with the internal dynamics of the system. Naturally, this approximation is restrictive and cannot be applied to arbitrary astrophysical systems. Therefore,  dynamical friction studies that deal with ordinary tidal interactions have been mostly based on $N$-body 
simulations~\cite{Munoz:2005be,Renaud,BSN, Wang:2016qol, Iorio:2019atc}. Indeed, various $N$-body simulations of Milky Way (MW) satellites have reported that the stellar components of these dwarf spheroidal galaxies are not directly affected by the tidal field of the MW~\cite{BSN, Wang:2016qol, Iorio:2019atc, DES:2018jtu}; that is, the influence of tidal effects on the stellar kinematics must be rather small. 

The main purpose of the present work is to extend Eq.~\eqref{I2} in the presence of tidal interactions. Specifically, we study the two-body Newtonian scattering of $M$ from a star of mass $m$ within a stellar system that is under the tidal influence of a distant mass. To simplify matters, we expand the solution of equation of relative motion of the binary in powers of the Newtonian gravitational constant and work to second order in $G$, since dynamical friction first appears at this order in accordance with Eq.~\eqref{I2}. In Section II, we derive the equation of relative motion in the presence of the tides. To solve this equation, we present a perturbation scheme in powers of $G$ in Section III and estimate the influence of tides on dynamical friction. The astrophysical implications of our results are briefly treated in Section IV. Finally, we discuss the limitations of our approach in Section V.

\section{Tidal Perturbation of the Gravitational Two-Body System}

Imagine a Newtonian two-body system with masses $m_1$ and $m_2$ within a stellar medium (of extent $D$) that is tidally perturbed by a distant galactic mass $\mathbb{M}$, $\mathbb{M} \gg m_1+m_2$. We focus attention on the two-body system and neglect the gravitational interaction of the two bodies with the stars of the background medium. The external source  $\mathbb{M}$ generates a gravitational potential $\Phi$ at the location of the two-body system. The equations of motion of $m_1$ and $m_2$ are given in a local Cartesian coordinate system by
\begin{equation}\label{T1} 
m_1\,\frac{d^2\mathbf{x}_{m_1}}{dt^2}= - \frac{Gm_1m_2 \mathbf{x}}{r^3} -m_1\,(\nabla \Phi)_1\, 
\end{equation}
and
\begin{equation}\label{T2} 
m_2\,\frac{d^2\mathbf{x}_{m_2}}{dt^2}=  \frac{Gm_1m_2 \mathbf{x}}{r^3} -m_2\,(\nabla \Phi)_2\,,
\end{equation}
where
\begin{equation}\label{T3} 
 \mathbf{x}:= \mathbf{x}_{m_1} - \mathbf{x}_{m_2}\,, \qquad r := |\mathbf{x}|\,.  
\end{equation}
We assume that $r \le D \ll \mathbb{R}$, where $\mathbb{R}$ is the distance of the binary system to $\mathbb{M}$. We expand the gravitational force of $\mathbb{M}$ on the two-body system about its center of mass
\begin{equation}\label{T4} 
(\nabla \Phi)_1^i = (\nabla \Phi)_{\rm cm}^i + K^{i}{}_{j}(\mathbf{x}_{\rm cm}) (x_{m_1}^j - x_{\rm cm}^j) + \cdots\, 
\end{equation}
and
\begin{equation}\label{T5} 
(\nabla \Phi)_2^i = (\nabla \Phi)_{\rm cm}^i + K^{i}{}_{j}(\mathbf{x}_{\rm cm}) (x_{m_2}^j - x_{\rm cm}^j) + \cdots\,, 
\end{equation}
where $K_{ij} = \nabla_i \nabla_j \Phi$ is the symmetric and traceless tidal matrix evaluated at the center of mass of the binary system
\begin{equation}\label{T6} 
\mathbf{x}_{\rm cm}= \frac{m_1\mathbf{x}_{m_1} +m_2 \mathbf{x}_{m_2}}{m_1+m_2}\,. 
\end{equation}
We assume that $(D/\mathbb{R})^2$ is negligibly small compared to unity; therefore, we henceforth neglect terms of  second and higher orders in Eqs.~\eqref{T4} and~\eqref{T5}. Then, the equations of motion imply
\begin{equation}\label{T7} 
\frac{d^2\mathbf{x}_{\rm cm}}{dt^2} = - (\nabla \Phi)_{\rm cm}\, 
\end{equation}
and 
\begin{equation}\label{T8} 
\frac{d^2 x^i}{dt^2} = - \frac{G(m_1+ m_2) x^i}{r^3} - K^{i}{}_{j}(\mathbf{x}_{\rm cm})\,x^j\,.  
\end{equation}
Appendix A contains the natural extension of the two-body approach developed here to an $N$-body system. However, it is not known how to derive the dynamical friction force within this scheme; therefore, in conformity with Chandrasekhar's approach, we hereafter ignore the gravitational interaction between the background stars. 

To interpret Eqs.~\eqref{T7} and~\eqref{T8}, we note that our dynamical system has a slow motion given by Eq.~\eqref{T7} and a fast motion given by Eq.~\eqref{T8}. The slow motion involves the Keplerian revolution of the whole binary system about $\mathbb{M}$. The fast motion involves the internal scattering orbit of the tidally perturbed gravitational two-body system. We assume that during the fast motion,  the slow motion can be considered to be essentially uniform. That is, the deviation of Keplerian orbit of the center of mass of the binary from a straight line with constant velocity can be neglected during the fast motion of the binary. 
This approximation scheme can be naturally extended to include the background stellar medium in order to facilitate the calculation of dynamical friction.  Moreover, in Eq.~\eqref{T8}, we neglect the temporal dependence of the tidal matrix $K^{i}{}_{j}(\mathbf{x}_{\rm cm})$ in our approach. In general, the temporal variation of the tidal field can lead to the possibility of resonance with the internal pulsation of the system; in fact,  such a resonant orbital coupling has been investigated in~\cite{Kuhn1, Kuhn2}. 

In the  inertial reference frame that moves with the constant velocity of the center of mass, mass $m_1$ with state $(\mathbf{x}_{m_1}, \mathbf{v}_{m_1})$ interacts gravitationally with mass $m_2$ with state $(\mathbf{x}_{m_2}, \mathbf{v}_{m_2})$ and the relative motion in given by Eq.~\eqref{T8}. The solution of this equation and the scattering process will be described in detail in the next section. However, it is simple to conclude from the uniform motion of the center of mass and  the definition of relative motion that the net change in the velocities of $m_1$ and $m_2$ are given by $\delta \mathbf{v}_{m_1} = m_2 \delta \mathbf{v}/(m_1+m_2)$ and $\delta \mathbf{v}_{m_2} = - m_1 \delta \mathbf{v}/(m_1+ m_2)$, where $\mathbf{v}:=\dot{\mathbf{x}}$ is the relative velocity.  

It remains to evaluate the external force acting on the center of mass in Eq.~\eqref{T7} and the tidal matrix in Eq.~\eqref{T8}. Let $\mathbb{R}\,\mathbf{n}$ be the vector that connects $\mathbb{M}$ to the center of mass of the binary system and $\mathbf{n}$ be the corresponding unit vector. Then, 
\begin{equation}\label{T9} 
(\nabla \Phi)_{\rm cm} = \frac{G\,\mathbb{M}}{\mathbb{R}^2}\,\mathbf{n}\,  
\end{equation}
and
\begin{equation}\label{T10} 
K^{i}{}_{j}(\mathbf{x}_{\rm cm}) = \frac{G\,\mathbb{M}}{\mathbb{R}^3}\,(\delta^i_j - 3 n^i n_j)\,.  
\end{equation}
In the local Cartesian coordinate system $\mathbf{x} = (x, y, z)$ established in the inertial frame that moves with constant velocity of the center of mass of the binary, we can choose the unit vector $\mathbf{n}$ to point in the $z$ direction with no loss in generality. The equation for relative motion of the binary~\eqref{T8} now takes the form
\begin{equation}\label{T11} 
\frac{d^2\mathbf{x}}{dt^2} = -G\left(\frac{M'}{r^3} + \lambda\right)\mathbf{x} + 3G\lambda (\mathbf{x}\cdot \mathbf{n})\mathbf{n}\,,  
\end{equation}
where we have introduced
\begin{equation}\label{T12} 
m_1 + m_2 := M'\,, \qquad \lambda := \frac{\mathbb{M}}{\mathbb{R}^3}\,.  
\end{equation}

Inspection of equation of relative motion of the binary~\eqref{T11} reveals that the external tidal force becomes comparable to the internal Newtonian inverse-square force at $r \approx R_0$, where $R_0$ is the \emph{tidal radius} given by
\begin{equation}\label{T13} 
\frac{M'}{R_0^3} = \frac{\mathbb{M}}{\mathbb{R}^3}\,, \qquad  R_0 = \left(\frac{M'}{\lambda}\right)^{1/3}\,.  
\end{equation}
Indeed, for $r > R_0$ the tidal force is dominant, while for $r < R_0$ the Newtonian attractive force between $m_1$ and $m_2$ is dominant. The main purpose of the present paper is to determine how the tidal interaction modifies the dynamical friction force that is a consequence of gravitational drag as $m_1$ scatters from $m_2$. Therefore, we must limit our considerations to regions within a sphere of radius 
$R_0$. In such regions, the tidal force is smaller than the internal Newtonian force of attraction and can be treated as a \emph{linear perturbation} on the binary. This approach is consistent with the limiting situation in which the exterior tidal force disappears; that is, for $\lambda \to 0$, we find $R_0 \to \infty$ and our treatment of dynamical friction reduces to the standard method~\cite{Chandrasekhar:1943ys,BT, TW}. We treat the influence of tides  as a first-order perturbation; therefore, we expect that the external tidal interaction will have a generally small effect on the internal dynamical friction force. The main purpose of this work is to determine the extent to which dynamical friction is modified by external tides. 

\section{Dynamical Friction in the Presence of Tides}

To calculate the dynamical friction force in the presence of tides, we seek a perturbative solution of the equation of relative motion~\eqref{T11} within the domain bounded by the tidal radius $R_0$. Let us first observe that Eq.~\eqref{T11} is invariant under a constant temporal translation as well as a constant rotation in the $(x, y)$ plane. We use these symmetries below.

\subsection{Initial Motion in the $(x, y)$ Plane}
\label{sx}

Let us first assume for the sake of simplicity that the unperturbed uniform relative motion occurs in the $(x, y)$ plane. We can therefore rotate the Cartesian coordinate system about the $z$ axis such that the unperturbed relative motion occurs along the $x$ axis from $(-x_0, y_0, z_0)$ to $(x_0, y_0, z_0)$. Here, $x_0 = v_0\, t_0$, $y_0$ and $z_0$ are constants that characterize the unperturbed motion. That is, 
\begin{equation}\label{F1} 
x(t) = v_0\,t, ~y = y_0, ~z = z_0\,, \qquad t = - t_0 \to t = t_0\,  
\end{equation}
such that 
\begin{equation}\label{F2} 
r_0 = (v_0^2\, t_0^2 + y_0^2 + z_0^2)^{1/2} < R_0\,.  
\end{equation}
Let us note here that we have used the invariance under time translation to set $x(t = 0) = 0$. Henceforth, we will express the unperturbed motion in the background $(x, y, z)$ coordinate system in the form 
\begin{equation}\label{F3} 
\boldsymbol{\xi}(t) = \mathbf{b} + \mathbf{v}_0\,t\,,   \qquad \mathbf{b} = (0, y_0, z_0)\,, \qquad \mathbf{v}_0 = (v_0, 0, 0)\,, 
\end{equation}
where $\mathbf{b} \cdot \mathbf{v}_0 = 0$ and $|\mathbf{b}| = b$ is the impact parameter. 

Let us recall here that dynamical friction is a phenomenon that first appears at order $G^2$ in the standard treatment~\cite{Chandrasekhar:1943ys, BT}. We therefore solve Eq.~\eqref{T11} perturbatively to second order in the gravitational constant $G$ and to linear order in the tidal parameter $\lambda$. Specifically, we assume a solution of the form
\begin{equation}\label{F4} 
\mathbf{x} (t) = \boldsymbol{\xi}(t)  +G\,\mathbf{x}_1(t) + G^2\,\mathbf{x}_2(t) + \cdots\,,  
\end{equation}
where we have treated $G$ as the expansion parameter. Similarly, we write
\begin{equation}\label{F5} 
\mathbf{v} = \frac{d\mathbf{x}}{dt}  = \mathbf{v}_0 + G\,\frac{d\mathbf{x}_1}{dt} + G^2\,\frac{d\mathbf{x}_2}{dt} + \cdots\,.  
\end{equation}
Henceforth, we drop terms of order $G^3$ and higher. To proceed, it is useful to introduce the quantities
\begin{equation}\label{F6} 
U(t)  :=  (\boldsymbol{\xi}\cdot \boldsymbol{\xi})^{1/2} = (v_0^2\, t^2 + b^2)^{1/2}\, 
\end{equation}
and
\begin{equation}\label{F7}
W(t) := \frac{\boldsymbol{\xi}\cdot \mathbf{x}_1}{U^2}\,, 
\end{equation}
in terms of which we can write
\begin{equation}\label{F8} 
\frac{1}{r^3} =  \frac{1}{U^3}\,(1 - 3 G W) + O(G^2)\,.  
\end{equation}

Plugging these expressions in Eq.~\eqref{T11}, we find to first order in $G$
\begin{equation}\label{F9} 
\frac{d^2 \mathbf{x}_1}{dt^2} = -  (\lambda + M'\,U^{-3})\boldsymbol{\xi}(t) + 3\lambda (\boldsymbol{\xi}\cdot \mathbf{n})\mathbf{n}\, 
\end{equation}
and to second order in $G$
\begin{equation}\label{F10} 
\frac{d^2 \mathbf{x}_2}{dt^2} = -  (\lambda + M'\,U^{-3})\mathbf{x}_1 + 3\lambda (\mathbf{x}_1\cdot \mathbf{n})\mathbf{n} + 3 M' U^{-3}W\boldsymbol{\xi}\,.  
\end{equation}
We must now solve these equations with the boundary conditions that initially at $t = -t_0$, we have $\mathbf{x}_1  = 0$, $\mathbf{x}_2  = 0$, $\dot{\mathbf{x}}_1 = 0$ and 
$\dot{\mathbf{x}}_2 = 0$. Here, $\dot{x}_1 := dx_1/dt$, etc.

Using standard integrals, given for convenience in Appendix B, we find 
\begin{equation}\label{F11} 
\dot{x}_1 = \frac{M'}{v_0}\,(U^{-1} - U_0^{-1}) - \frac{1}{2}\,\frac{\lambda}{v_0}\,(u^2 - u_0^2)\,,
\end{equation}
\begin{equation}\label{F12} 
\dot{y}_1 = -\frac{M'\,y_0}{v_0b^2}\,(u\,U^{-1} +u_0\, U_0^{-1}) - \frac{\lambda}{v_0}\,y_0\,(u+u_0)\,,
\end{equation}
\begin{equation}\label{F13} 
\dot{z}_1 = -\frac{M'\,z_0}{v_0b^2}\,(u\,U^{-1} +u_0\, U_0^{-1}) + 2 \frac{\lambda}{v_0}\,z_0\,(u+u_0)\,,
\end{equation}
where, for the sake of convenience,  we have introduced  
\begin{equation}\label{F14} 
 u := v_0\,t\,, \quad u_0 := x_0 = v_0\,t_0\,, \quad U_0 := r_0  =  (u_0^2 + b^2)^{1/2}\,.
\end{equation}
Let us note from Eq.~\eqref{F11} that $\dot{x}_1(t_0) = 0$ and hence $\delta \dot{x}_1 = 0$, which is the expected result at linear order in $G$ from the standard treatment of dynamical friction. In general, the situation is different in the presence of tides, as will be demonstrated in the last part of this section; however, the tidal contribution does vanish in certain special cases that include initial motions along the $x$ and $z$ axes. 

Integration of Eqs.~\eqref{F11}--\eqref{F13} with the appropriate initial conditions results in
\begin{equation}\label{F15} 
x_1 = \frac{M'}{v_0^2}\,\mathbb{A}_1 - \frac{1}{6}\, \frac{\lambda}{v_0^2}\,\mathbb{B}_1\,,
\end{equation}
\begin{equation}\label{F16} 
y_1 = -\frac{M'\,y_0}{v_0^2\,b^2}\,\mathbb{A}_2 - \frac{1}{2}\, \frac{\lambda}{v_0^2}\,y_0\,\mathbb{B}_2\,,
\end{equation}
\begin{equation}\label{F17} 
z_1 = -\frac{M'\,z_0}{v_0^2\,b^2}\,\mathbb{A}_2 +  \frac{\lambda}{v_0^2}\,z_0\,\mathbb{B}_2\,.
\end{equation}
Here, we have introduced, for the sake of convenience, the functions 
\begin{equation}\label{F17a} 
\mathbb{A}_1(u) := \ln\left(\frac{U + u}{U_0 - u_0}\right) - \frac{u +u_0}{U_0}\,, \qquad \mathbb{A}_2(u) := U - U_0 + \frac{u_0}{U_0}\,(u+u_0)\,
\end{equation}
and
\begin{equation}\label{F17b} 
\mathbb{B}_1(u) := u^3 - 3 u_0^2\,u - 2u_0^3\,, \qquad \mathbb{B}_2(u) := (u+u_0)^2\,.
\end{equation}

To go further, we can integrate the second-order perturbation equations for $(x_2, y_2, z_2)$. On the other hand, for the purpose of calculating dynamical friction we only need the equation for $x_2$; that is, we must integrate the first component of Eq.~\eqref{F10} from $t = -t_0$ to $t = t_0$.  More precisely, we need to calculate 
\begin{equation}\label{F18} 
\delta v_x = v_x(t_0) -v_x(-t_0) = v_x(t_0) - v_0 =  G\,\frac{dx_1}{dt}(t_0) + G^2\,\frac{dx_2}{dt}(t_0) = G^2\,\frac{dx_2}{dt}(t_0)\,,
\end{equation}
since we already know that $\dot{x}_1(t_0) = 0$, i.e. the term linear in $G$ vanishes in this case. We treat the tidal parameter 
$\lambda$ to first order in our perturbation analysis; therefore, after some algebra we find 
\begin{equation}\label{F19} 
\frac{d^2 x_2}{dt^2} =  \frac{M'^2}{v_0^2U_0U^5}\, \mathcal{A} +\frac{M'\,\lambda}{v_0^2}\,\left(\frac{\mathcal{B}_1}{U^5}  + \mathcal{B}_2\right) + O(\lambda^2)\,.
\end{equation}
Here, 
\begin{equation}\label{F20} 
\mathcal{A} = U_0[(2u^2-b^2)\,\mathbb{A}_1 - 3u\,\mathbb{A}_2]\,,
\end{equation}
\begin{equation}\label{F21} 
\mathcal{B}_1 = -\frac{1}{6}(2u^2-b^2)\,\mathbb{B}_1 +\frac{3}{2}(2z_0^2-y_0^2) u\,\mathbb{B}_2\,,
\end{equation}
\begin{equation}\label{F22} 
\mathcal{B}_2 = - \mathbb{A}_1\,.
\end{equation}

Next, we integrate Eq.~\eqref{F19} from $t = - t_0$ to $t = t_0$ with the boundary condition that $\dot{x}_2(- t_0) = 0$. The result is 
\begin{equation}\label{F23} 
\dot{x}_2( t_0) =  \frac{2M'^2u_0}{v_0^3U_0^3}\,\mathcal{S} +  \frac{2M'\,\lambda \,u_0}{v_0^3}\mathcal{T}\,,
\end{equation}
where
\begin{equation}\label{F24} 
\mathcal{S}  =  \mathbb{C} - \frac{u_0^3}{b^2U_0}\,, \qquad \mathcal{T}  =  \mathbb{C} + \frac{u_0^3}{3b^2U_0^3}(5z_0^2-4y_0^2)\,.
\end{equation}
Here, $\mathbb{C}(b, u_0) \le 0$ is given by
\begin{equation}\label{F25} 
 \mathbb{C}(b, u_0)  :=  \ln\left(\frac{U_0-u_0}{b}\right)  +\frac{u_0}{U_0}\,.
\end{equation}

We can now derive the modification of Eq.~\eqref{I2} in the presence of tides to second order in $G$. Suppose that $m_1 = M$, $m_2 = m$ and $M + m = M'$. Furthermore, our perturbative  approach implies
\begin{equation}\label{F26} 
(\delta \mathbf{v}_M)_{||} = (m/M') (\delta \mathbf{v})_{||} = (m/M')\frac{\mathbf{v}_0}{v_0} G^2 \dot{x}_2( t_0)\,.
\end{equation}
Using Eq.~\eqref{F23}, we find
\begin{equation}\label{F27} 
(\delta \mathbf{v}_M)_{||} = 2G^2m\,\frac{\mathbf{v}_0}{v_0^4}\left[\frac{M'\,u_0}{U_0^3}\,\mathcal{S} + \lambda \,u_0\mathcal{T}\right]\,,
\end{equation}
which is the analogue of Eq.~\eqref{I2} in the presence of tides. We note here the remarkable fact that $\mathcal{S}$ is negative; more specifically, $\mathcal{S}$ as a function of $u_0/b$ starts out from zero  at $u_0/b = 0$ with vanishing slope and then monotonically decreases to $- \infty$ as $u_0/b \to \infty$. Moreover, let us note that in the presence of tides $t_0 > 0$ is  finite and is limited by Eq.~\eqref{F2}; however, in the absence of tides $\lambda = 0$ and we can let $t_0 \to \infty$ as in the standard treatment~\cite{BT}. Therefore, in the absence of tides, $t_0 \to \infty$,  $u_0 = v_0\,t_0 \to \infty$ and  $\mathcal{S} \to -u_0^2/b^2$. Then, Eq.~\eqref{F27} reduces to
\begin{equation}\label{F28} 
(\delta \mathbf{v}_M)_{||} = -\frac{2G^2 m (m+M)\mathbf{v}_0}{b^2\,v_0^4}\,,
\end{equation}
in agreement with Eq.~\eqref{I2} to second order in $G$. 

A comment is in order here regarding the fact that $t_0 = u_0/v_0$ is essentially a free parameter in Eq.~\eqref{F27}.  Indeed, in the  application of Chandrasekhar's formula~\eqref{I1} to actual astrophysical systems, the scattering of $M$ from each star of mass $m$ in the system does involve a different finite $t_0$ that we simply ignore. Instead, in each such case, we employ the result obtained from the ideal case with $t_0 = \infty$. This approximation does not ordinarily encounter any obstacles. However, the presence of a finite tidal radius $R_0$ in the case under consideration implies that $t_0$ is a free parameter subject to the restriction contained in the inequality $v_0^2\,t_0^2 + b^2  < R_0^2$.

To find the analogue of Chandrasekhar's formula~\eqref{I1} in the present case, we need to extend Eq.~\eqref{F27} for an arbitrary star $m$ to the background stars by taking the flux of the stars into account  and integrating over the appropriate state space of the background stars. Let  $f(\mathbf{x}_m, \mathbf{v}_m, t)$ be the state-space number density of background stars. In the idealized infinite homogeneous background medium employed in the original derivation of Eq.~\eqref{I1}, the distribution function $f$ reduces to $f_0$, which is independent of temporal and spatial variables and is an isotropic function of $\mathbf{v}_m$; however, in the presence of tides, the situation is quite different. Starting from an initial distribution function, it is possible to take the time dependence of tides into account and then basically integrate the tidal force (per unit mass of a star) over a certain interval of time and in this way come up with the change in the velocity of the star due to the corresponding tidal impulse. In principle, the evolution of the distribution function due to the presence of tides can thereby be determined~\cite{BT, spitzer, gnedin, Banik}. The determination of the appropriate distribution function in the case under consideration requires a separate investigation and is beyond the scope of the present work. 

To get a rough estimate of the influence of tides on dynamical friction, we henceforth ignore the difference between $f(\mathbf{x}_{m}, \mathbf{v}_{m}, t)$ and $f_0(v_{m})$. The net rate of change of momentum $M \mathbf{v}_M$ per unit time is then given by 
\begin{equation}\label{F28a} 
M\frac{d\mathbf{v}_{M}}{dt} \approx 2 G^2 mM \int \left[\frac{M'\,u_0}{U_0^3}\,\mathcal{S} +  \lambda \,u_0\mathcal{T}\right] dy_0\,dz_0 \int \frac{\mathbf{v}_{M} - \mathbf{v}_{m}}{|\mathbf{v}_{M} - \mathbf{v}_{m}|^3} f_0(v_m)d^3v_m\,,
\end{equation}
where $\mathbf{v}_0$ has been replaced by $\mathbf{v}_{M} - \mathbf{v}_{m}$, as in the standard procedure~\cite{BT}. For an isotropic function $f_0$, Newton's shell theorem implies~\cite{BT}
\begin{equation}\label{F28b} 
\int \frac{\mathbf{v}_{M} - \mathbf{v}_{m}}{|\mathbf{v}_{M} - \mathbf{v}_{m}|^3} f_0(v_m)d^3v_m = 4\pi \frac{\mathbf{v}_M}{v_M^3} \int_0^{v_M} f_0(v_m) v_m^2\, dv_m\,.
\end{equation}
Using
\begin{equation}\label{F28c} 
\rho(< v_M)   = 4\pi m \int_0^{v_M} f_0(v_m) v_m^2\, dv_m\,,
\end{equation}
we finally arrive at 
\begin{equation}\label{F29} 
M\frac{d\mathbf{v}_{M}}{dt} \approx 2 G^2 M\frac{\mathbf{v}_M}{v_M^3}\,\rho(< v_M) \int \left[\frac{M'\,u_0}{U_0^3}\,\mathcal{S} +  \lambda \,u_0\mathcal{T}\right] dy_0\,dz_0\,.
\end{equation}

Let us proceed with the integration over $dy_0\wedge dz_0$. Introducing the azimuthal angle $\varphi$, we have
\begin{equation}\label{F30} 
y_0 = b\,\cos\varphi\,, \qquad z_0 = b\,\sin\varphi\,.
\end{equation}
Then, $dy_0\wedge dz_0 = b\,db \wedge d\varphi$  and one can show that 
\begin{equation}\label{F31} 
 \int_{b_{\rm min}}^{b_{\rm max}} \frac{u_0\,\mathcal{S}}{U_0^3}\, b\,db \,d\varphi = \mathbb{S}(b_{\rm max}, u_0|_{b_{\rm max}}) - \mathbb{S}(b_{\rm min}, u_0|_{b_{\rm min}})\,,
\end{equation}
where
\begin{equation}\label{F32} 
\mathbb{S}(b, u_0) := -\frac{2\pi u_0}{U_0}\, \mathbb{C}(b, u_0)\,.
\end{equation}
Let us observe that $\mathbb{S}(b, u_0)$ vanishes for $u_0=0$, while for $u_0 > 0$, $\mathbb{S}(b, u_0)$  is a positive function of $b/u_0$ that monotonically decreases from $\infty$ at $b/u_0 = 0$ and tends to zero as $b/u_0 \to \infty$. 

In the absence of tides, $\lambda = 0$, $u_0 \to \infty$ and $\mathcal{S} \to -u_0^2/b^2$; then, as expected,  we recover $-\ln (b_{\rm max}/b_{\rm min})$ from Eq.~\eqref{F31}. However, in the presence of tides $u_0^2 + b^2 = r_0^2 < R_0^2$ according to Eq.~\eqref{F2}; therefore, in Eq.~\eqref{F31}, $\mathbb{S}(b_{\rm max}, u_0|_{b_{\rm max}})$ vanishes, since $b_{\rm max} = r_0$ and hence $u_0|_{b_{\rm max}}=0$. To interpret this result physically, let us first note that $u_0 = v_0 \,t_0$ vanishes when either $v_0 = 0$ or $t_0 = 0$. It is clear from Eq.~\eqref{I2} that if $v_0 = 0$, there is no scattering and dynamical friction force vanishes. On the other hand, if $t_0$ vanishes, there is no time for interaction and hence no change in momentum is possible. Thus, for the nontidal part of the integral in Eq.~\eqref{F29}, we find  
\begin{equation}\label{F33} 
\int \frac{M'\,u_0}{U_0^3}\,\mathcal{S}\, dy_0\,dz_0 = -(m+M)\,\mathbb{S}(b_{\rm min}, u_0|_{b_{\rm min}})\,, \qquad b_{\rm min}^2 +u_0|_{b_{\rm min}}^2 = r_0^2\,.
\end{equation}
In our case, the nontidal part is thus $-(m+M)$ times the analogue of the Coulomb logarithm. As in the standard treatment of dynamical friction~\cite{BT}, we assume $b_{\rm min}$ is the maximum of the size of the incident mass $M$ and $GM/v_M^2$, while $r_0 = \min(D,R_0)$, where $D$ is the diameter of the galactic system. In the absence of tides, the tidal radius $R_0$ goes to infinity and consequently we have $r_0=D$. 

\begin{figure}
\centering
\includegraphics[width=8.0cm]{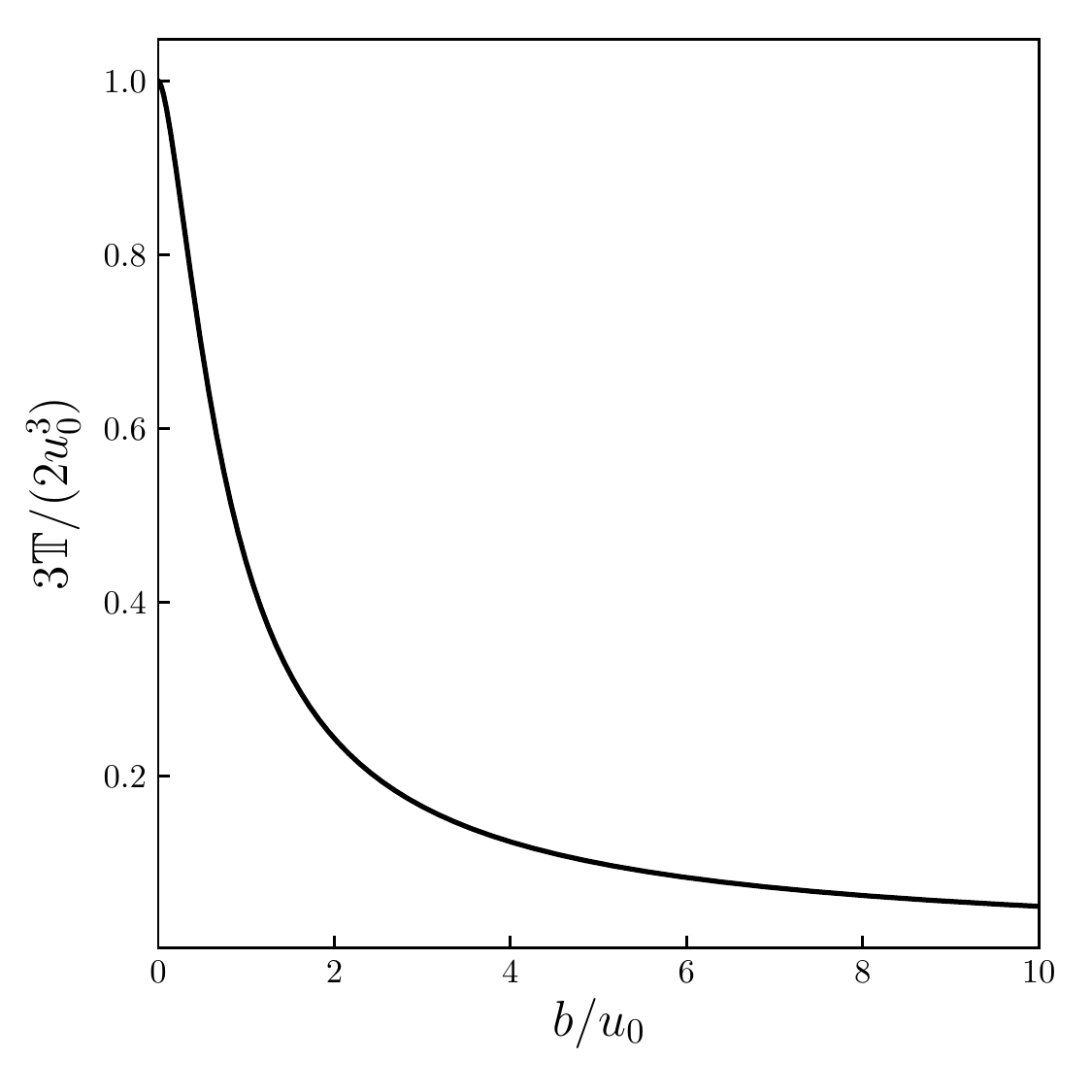}
\caption{Plot of the tidal contribution to dynamical friction for motion along the $x$ axis.}\label{fig1}
\end{figure}

Let us next consider the tidal part of Eq.~\eqref{F29}, namely, the integral of $\lambda\,u_0\,\mathcal{T}$ that turns out to be 
\begin{equation}\label{F34} 
\lambda\, \int_{b_{\rm min}}^{b_{\rm max}} u_0\,\mathcal{T} b\,db \,d\varphi = \pi\,\lambda [\mathbb{T}(b_{\rm max}, u_0|_{b_{\rm max}}) - \mathbb{T}(b_{\rm min}, u_0|_{b_{\rm min}})]\,,
\end{equation}
where $\mathbb{T}$ is proportional to $u_0$ and is given by
\begin{equation}\label{F35} 
\mathbb{T}(b, u_0)  := b^2\,u_0\left(\mathbb{C} + \frac{2}{3}\frac{u_0^3}{b^2U_0}\right)\,.
\end{equation}
As illustrated in Figure~\ref{fig1}, the function $3 \mathbb{T}(b, u_0)/(2u_0^3)$ versus $b/u_0$ starts from unity with zero slope at $b/u_0 = 0$ and monotonically decreases to zero as $b/u_0 \to \infty$.
As noted before, $u_0|_{b_{\rm max}}=0$ implies that $\mathbb{T}(b_{\rm max}, u_0|_{b_{\rm max}}) = 0$; therefore, 
\begin{equation}\label{F36} 
\lambda\, \int_{b_{\rm min}}^{b_{\rm max}} u_0\,\mathcal{T} b\,db \,d\varphi =  -\pi\,\lambda \, \mathbb{T}(b_{\rm min}, u_0|_{b_{\rm min}})\,.
\end{equation}

Finally, to determine the relative strength of the tidal part compared to the nontidal part in the calculation of dynamical friction force in the case of motion along the $x$ axis, we must calculate $\Psi_x$,
\begin{equation}\label{F37} 
\Psi_x = \frac{\pi\,\lambda}{M'}\frac{\mathbb{T}(b_{\rm min}, u_0|_{b_{\rm min}})}{\mathbb{S}(b_{\rm min}, u_0|_{b_{\rm min}})}\,, \qquad  R_0 = \left(\frac{M'}{\lambda}\right)^{1/3}\,,
\end{equation}
once we ignore the influence of tides on the distribution function of the stars. In Section IV, we estimate $\Psi_x$ for the case of Fornax dSph galaxy tidally perturbed by the Galaxy (MW). 

\subsection{Initial Motion Along the $z$ Axis}

Let us assume that the unperturbed motion is given by
\begin{equation}\label{Z1} 
x = \bar{x}_0\,, ~y = \bar{y}_0\,, ~z = v_0\,t\,,  \qquad t = - t_0 \to t = t_0\,, \qquad  \bar{x}_0^2 + \bar{y}_0^2 = b^2\,
\end{equation}
such that $r_0 = U_0 =  (u_0^2 + b^2)^{1/2} < R_0$ as before. That is, 
\begin{equation}\label{Z2} 
\boldsymbol{\xi}(t) = \mathbf{b} + \mathbf{v}_0\,t\,,   \qquad \mathbf{b} = (\bar{x}_0, \bar{y}_0, 0)\,, \qquad \mathbf{v}_0 = (0, 0, v_0)\,.
\end{equation}
Let us note here that we have used the invariance under time translation to set $z(t =0) = 0$. 

Following the same procedures as in the previous subsection, we can obtain $\mathbf{x}_1$ and $\mathbf{x}_2$ for the perturbed motion. In the case of the linear deviation $\mathbf{x}_1$, inspection of Eq.~\eqref{T11} reveals that we can find the results in this case from Eqs.~\eqref{F15}--\eqref{F17} for motion along the $x$ direction by simply letting $\lambda \to -2\,\lambda$. Indeed, 
\begin{equation}\label{Z3} 
x_1 = -\frac{M'\,\bar{x}_0}{v_0^2\,b^2}\,\mathbb{A}_2 - \frac{1}{2}\,\frac{\lambda}{v_0^2}\,\bar{x}_0\,\mathbb{B}_2\,,
\end{equation}
\begin{equation}\label{Z4} 
y_1 = -\frac{M'\,\bar{y}_0}{v_0^2\,b^2}\,\mathbb{A}_2  - \frac{1}{2}\, \frac{\lambda}{v_0^2}\,\bar{y}_0\,\mathbb{B}_2\,,
\end{equation}
\begin{equation}\label{Z5} 
z_1 = \frac{M'}{v_0^2}\,\mathbb{A}_1  + \frac{1}{3}\,\frac{\lambda}{v_0^2}\,\mathbb{B}_1\,.
\end{equation}

We are interested in the third component of Eq.~\eqref{F10} in the present case. After some straightforward calculations, we find
\begin{equation}\label{Z6} 
\frac{d^2 z_2}{dt^2} =  \frac{M'^2}{v_0^2U_0U^5}\, \bar{\mathcal{A}} +\frac{M'\,\lambda}{v_0^2}\,\left(\frac{\bar{\mathcal{B}}_1}{U^5}  + \bar{\mathcal{B}}_2\right) + O(\lambda^2)\,.
\end{equation}
Here, the nontidal part is the same as in the case of motion along the $x$ direction by the spherical symmetry of the Newtonian gravitational potential of a point mass; that is,
\begin{equation}\label{Z7} 
\bar{\mathcal{A}} = \mathcal{A}\,, 
\end{equation}
while changes occur in the tidal part. Indeed, 
\begin{equation}\label{Z8} 
\bar{\mathcal{B}}_1 = \frac{1}{3}(2u^2-b^2)\mathbb{B}_1 -\frac{3}{2}\,b^2 u\,\mathbb{B}_2\,, \quad \bar{\mathcal{B}}_2 = - 2\,\mathcal{B}_2\,. 
\end{equation}
It follows that in this case
\begin{equation}\label{Z9} 
(\delta \mathbf{v}_{M})_{||} = 2G^2m\,\frac{\mathbf{v}_0}{v_0^4}\left[\frac{M'\,u_0}{U_0^3}\,\mathcal{S} +  \lambda \,u_0\bar{\mathcal{T}}\right]\,,
\end{equation}
where
\begin{equation}\label{Z10} 
\bar{\mathcal{T}}  = -2\,\mathbb{C} - \frac{u_0^3}{3U_0^3}\,.
\end{equation}
Following essentially the same arguments as before leads to the next step which involves 
\begin{equation}\label{Z11} 
M\frac{d\mathbf{v}_{M}}{dt} \approx 2 G^2 M\frac{\mathbf{v}_M}{v_M^3}\,\rho(< v_M) \int \left[\frac{M'\,u_0}{U_0^3}\,\mathcal{S} +  \lambda \,u_0\bar{\mathcal{T}}\right] d\bar{x}_0\,d\bar{y}_0\,.
\end{equation}
With $\bar{x}_0 = b\,\cos\bar{\varphi}$, $\bar{y}_0 = b\,\sin\bar{\varphi}$ and $d\bar{x}_0 \wedge d\bar{y}_0 = b\,db \wedge d\bar{\varphi}$, we find
\begin{equation}\label{Z12} 
\lambda\,\int_{b_{\rm min}}^{b_{\rm max}} u_0\,\bar{\mathcal{T}} b\,db \,d\bar{\varphi} = \pi\,\lambda [\bar{\mathbb{T}}(b_{\rm max}, u_0|_{b_{\rm max}}) - \bar{\mathbb{T}}(b_{\rm min}, u_0|_{b_{\rm min}})]\,.
\end{equation}
Here,  $\bar{\mathbb{T}}$ turns out to be
\begin{equation}\label{Z13} 
\bar{\mathbb{T}}(b, u_0)  = - 2\, \mathbb{T}(b, u_0)\,.
\end{equation}
The negative sign in Eq.~\eqref{Z13} means that the tidal stretching along the $z$ direction is clearly an antifriction force that opposes the dynamical friction force. 

Finally,  the relative strength of the tidal part compared to the nontidal part for motion along the $z$ direction is given by $\Psi_z = - 2 \Psi_x$, which we will estimate for the tidal perturbation of Fornax dSph galaxy by the MW in Section IV.

\subsection{General Treatment}

In the general case, the initial fixed relative velocity vector is arbitrary; however, the freedom in a constant rotation about the $z$ axis may be employed to render $\mathbf{v}_0$ in the $(x, z)$ plane with no loss in generality. As before, we can write the \emph{unperturbed} motion in the form 
\begin{equation}\label{W1} 
\boldsymbol{\xi}(t) = \mathbf{b} + \mathbf{v}_0\,t\,,  \qquad \mathbf{b} \cdot \mathbf{v}_0 = 0\,, 
\end{equation}
where  
\begin{equation}\label{W2} 
 \mathbf{v}_0 = v_0\,(\sin\theta, 0, \cos\theta)\,,  \qquad \mathbf{b} =  b\,(-\cos \theta \sin\phi, \cos \phi, \sin \theta \sin\phi)\,. 
\end{equation}
That is, $\mathbf{b} = b\,(\cos\phi \,\mathbf{e}_1 + \sin\phi\, \mathbf{e}_2)$, where $\mathbf{e}_1= (0, 1, 0)$ is the unit vector in the $y$ direction and $\mathbf{e}_2= (-\cos\theta, 0, \sin\theta)$ is the unit vector in the $(x, z)$ plane normal to $\mathbf{v}_0$. Here, $\theta: 0 \to \pi$ is the \emph{fixed} polar angle of the initial relative velocity vector, while the azimuthal angle $\phi: 0 \to 2\pi$ indicates the range of the possible directions of the impact parameter.  As before, the motion takes place from 
$t = - t_0 \to t = t_0$ and the freedom in time translation has been employed such that the midpoint of the unperturbed motion is at $t = 0$. Let us note that we work within the region bounded by the tidal radius $R_0$, namely, 
\begin{equation}\label{W3} 
r_0 =  [\boldsymbol{\xi}(\pm t_0)\cdot \boldsymbol{\xi}(\pm t_0)]^{1/2} = (v_0^2\, t_0^2 + b^2)^{1/2} < R_0\,.  
\end{equation}
The general case considered here simply reduces to the two previous cases when $(\theta, \phi)$ are chosen appropriately. Indeed, for $\theta = \pi/2$ and $\phi = \varphi$ the unperturbed motion is along the $x$ direction with $y_0 = b \cos \varphi$ and $z_0 = b \sin \varphi$, while for 
$\theta = 0 $ and $\phi = \bar{\varphi} - \pi/2$ we recover the unperturbed motion along the $z$ axis with $\bar{x}_0 = b \cos \bar{\varphi}$ and $\bar{y}_0 = b \sin \bar{\varphi}$. 

We now use the general Eqs.~\eqref{F4}--\eqref{F10} and the results of Appendix B to integrate the perturbation equations. We find 
\begin{align}\label{W4} 
\nonumber \dot{\mathbf{x}}_1 = {}& -\frac{M'}{v_0b^2}\Big[\frac{u}{U}+\frac{u_0}{U_0}+\frac{\lambda\,b^2}{M'}(u+u_0)\Big]\mathbf{b}+\frac{M'}{v_0^2}\Big[\frac{1}{U}-\frac{1}{U_0}-\frac{\lambda}{2\,M'}(u^2-u_0^2)\Big]\mathbf{v}_0\\
&+3\,\frac{\lambda}{v_0}\Big[(u+u_0)\,b \sin \theta \sin \phi+\frac{1}{2}(u^2-u_0^2)\,\cos \theta\Big]\,\mathbf{n}\,,
\end{align}
where $\mathbf{n} = (0, 0, 1)$ is the unit vector along the $z$ axis defined in Eq.~\eqref{T9}. Let us note that  $\delta \dot{\mathbf{x}}_1 = \dot{\mathbf{x}}_1(t_0)- \dot{\mathbf{x}}_1(-t_0)= \dot{\mathbf{x}}_1(t_0)$ can be simply calculated from Eq.~\eqref{W4} and is such that
\begin{equation}\label{W5}
\delta \dot{\mathbf{x}}_1 \cdot \mathbf{v}_0 = 6\,\lambda\,b\, u_0\sin\theta \cos\theta \sin\phi\,.
\end{equation}
This quantity is in general nonzero in the presence of tides. It does vanish, however, for some spacial cases including $\theta = 0$, i.e.  for motion along the $z$ axis, and $\theta = \pi/2$, i.e.  for motion along the $x$ axis.
 
Integration of Eq.~\eqref{W4} with the appropriate initial conditions gives
\begin{equation}\label{W6}
\mathbf{x}_1 = \frac{M'}{v_0^3}\, \mathbb{A}_1\mathbf{v}_0 -\frac{M'}{v_0^2b^2}\mathbb{A}_2\, \mathbf{b}- \frac{\lambda}{6v_0^3}(\mathbb{B}_1\, \mathbf{v}_0 + 3v_0\mathbb{B}_2\, \mathbf{b} - 3v_0\mathbb{B}_3\,\mathbf{n})\,,
\end{equation}
where  $\mathbb{B}_3$ is given by
\begin{equation}\label{W7}
\mathbb{B}_3 := \cos\theta \,\mathbb{B}_1 + 3b \sin\theta\sin\phi\,\mathbb{B}_2\,.
\end{equation}

The net change in relative velocity is given by $\delta \mathbf{v} =G\,\dot{\mathbf{x}}_1(t_0)+ G^2\,\dot{\mathbf{x}}_2(t_0)$. We are interested in 
\begin{equation}\label{W8}
\frac{1}{v_0}\delta \mathbf{v}\cdot\mathbf{v}_0 = \frac{1}{v_0}[6 G \lambda b u_0 \sin\theta \cos\theta \sin\phi + G^2\,\dot{\mathbf{x}}_2(t_0)\cdot\mathbf{v}_0]\,,
\end{equation}
where the linear term in $G$ is given by Eq.~\eqref{W5}. To find the equation for $\mathbf{x}_2$, we must substitute Eq.~\eqref{W6} into Eq.~\eqref{F10}. After some algebra, the result that we need is given by 
\begin{equation}\label{W9}
\frac{d^2\mathbf{x}_2}{dt^2} \cdot \mathbf{v}_0 = \frac{M'^2}{v_0U_0U^5}\,\mathcal{A} - \frac{M'\lambda}{6v_0U^5}\, \mathbb{B}_4 + \frac{M'\lambda}{2v_0b}\,\mathbb{B}_5 + O(\lambda^2)\,.
\end{equation}
 Here, $\mathcal{A}$ is given by Eq.~\eqref{F20} and
\begin{equation}\label{W10}
\mathbb{B}_4 := 6U^5\,\mathbb{A}_1 +(2u^2-b^2)\,\mathbb{B}_1+ 9 u b^2 \,\mathbb{B}_2 - 9 u (u \cos\theta + b \sin\theta\sin\phi)\,\mathbb{B}_3\,,
\end{equation}
\begin{equation}\label{W11}
\mathbb{B}_5 := [6 b\cos\theta\,\mathbb{A}_1 - 6\sin\theta\sin\phi\,\mathbb{A}_2 - b U^{-3}\,\mathbb{B}_3]\cos\theta\,.
\end{equation}
Next, we integrate Eq.~\eqref{W9} from $t = -t_0$ to $t = t_0$ and keep only the linear terms in $\lambda$. The result is
\begin{equation}\label{W12} 
\delta\dot{\mathbf{x}}_2 \cdot \mathbf{v}_0 =  \frac{2M'^2u_0}{v_0^2U_0^3}\,\mathcal{S} +  \frac{2M'\,\lambda \,u_0}{v_0^2}\,\mathcal{T}'\,,
\end{equation}
where $\mathcal{S}$ is given by Eq.~\eqref{F24} and $\mathcal{T}'$ can be expressed as
\begin{equation}\label{W13}
\mathcal{T}' = P_2(\cos\theta)\,\bar{\mathcal{T}} + \frac{3b}{2u_0}\left[\bar{\mathcal{T}} - \frac{u_0^3(2u_0^2+3b^2)}{U_0^3b^2}\right]\sin\theta \cos\theta \sin\phi - \frac{3}{2}\frac{u_0^3}{U_0^3} \sin^2\theta \cos2\phi\,.
\end{equation}
Here, $P_2(x)$ is the Legendre polynomial of second degree~\cite{A+S}
\begin{equation}\label{W13a}
P_2(x) := \frac{1}{2} (3x^2-1)\,.
\end{equation}
In Eq.~\eqref{W13}, $\mathcal{T}'$ reduces to $\bar{\mathcal{T}}$ for motion along the $z$ axis ($\theta = 0$) and to $\mathcal{T}$ for motion along the $x$ axis ($\theta = \pi/2$). 

Let us recall that $\delta \mathbf{v}_{M}=  (m/M') \delta \mathbf{v}$; therefore, 
\begin{equation}\label{W14} 
(\delta \mathbf{v}_M)_{||} = 6Gm\,\frac{\mathbf{v}_0}{v_0^2}\frac{\lambda\,b\, u_0}{M'}\sin\theta \cos\theta \sin\phi + 2G^2m\,\frac{\mathbf{v}_0}{v_0^4}\left[\frac{M'\,u_0}{U_0^3}\,\mathcal{S} + \lambda \,u_0\mathcal{T'}\right]\,.
\end{equation}
Next,  we need to sum the contribution of all of the stars of the background medium. As before, we neglect the influence of tides on the distribution function. In summing over the flux of stars, the spatial integration is over $b\, db\wedge d\phi$, as before. When we integrate over the azimuthal angle $\phi$,   terms proportional to $\sin\phi$ and $\cos 2\phi$ in Eqs.~\eqref{W13} and~\eqref{W14}  vanish. As in the special cases we have considered above, we find
\begin{equation}\label{W15} 
M\frac{d\mathbf{v}_{M}}{dt} \approx 2 G^2 M\frac{\mathbf{v}_M}{v_M^3}\,\rho(< v_M) \left[M'\,\mathbb{S}(b, u_0) + \pi  \lambda \,P_2(\cos\theta)\,\bar{\mathbb{T}}(b, u_0)\right]_{b_{\rm min}}^{b_{\rm max}}\,,
\end{equation}
or, finally, 
\begin{equation}\label{W16} 
M\frac{d\mathbf{v}_{M}}{dt} \approx  -2 G^2 M\frac{\mathbf{v}_M}{v_M^3}\,\rho(< v_M) \left[M'\,\mathbb{S}(b_{\rm min}, u_0|_{b_{\rm min}}) + \pi  \lambda \,P_2(\cos\theta)\,\bar{\mathbb{T}}(b_{\rm min}, u_0|_{b_{\rm min}})\right]\,,
\end{equation}
which reduces to our previous results for $\theta = \pi/2$ and $\theta = 0$. It is interesting to note that the tidal contribution to dynamical friction in our calculation vanishes for $P_2(\cos\theta) = 0$, which occurs for $\cos^2\theta = 1/3$, i.e. when the polar angle of the initial velocity is $\theta = \theta_0$ or $\theta = \pi - \theta_0$, where $\theta_0 \approx 54.7^\circ$.

In principle, the dynamical friction force can be used to distinguish between Newtonian gravitation with particle dark matter and the Newtonian limit of nonlocal gravity theory, where the nonlocal aspect of the gravitational interaction simulates dark matter. Consequently, within the context of nonlocal gravity, there is no particle dark matter halo around galaxies. On the other hand, in the standard dark matter picture, the main contribution to dynamical friction experienced by the galactic bars comes from the dark matter halo. Therefore, it is natural to expect different amounts of dynamical friction within the galactic systems modeled in the particle dark matter scenario or nonlocal gravity. Appendix C contains a brief discussion of how Eq.~\eqref{W16} can be extended to the Newtonian regime of nonlocal gravity.

\section{Astrophysical Implications}

As an application of our approach, we consider the Fornax dwarf spheroidal (dSph) galaxy, which is one the most luminous and widely studied dSph galaxies of the Local Group (LG). It is generally assumed that  the tidal forces of the MW have not had much impact on the stellar kinematics of Fornax; however, see~\cite{Hammer:2018bkq, Hammer:2020qcd}. In any case, there is no direct observational evidence for tidal disturbance of the stellar component of Fornax~\cite{Wang:2016qol}. However, tidal stripping can still significantly affect the more extended dark matter component of Fornax. It is necessary to mention that the dark matter content of Fornax inferred from its stellar kinematics is unexpectedly low, by a factor of $\sim 3$, compared to corresponding galaxies in cosmological simulations~\cite{Read:2018fxs}. Therefore, tidal stripping of its dark matter halo has been proposed as a solution to this discrepancy~\cite{Penarrubia:2007zx, GRF}.

 Another important feature of Fornax that is relevant to our discussion is the puzzling spatial distribution of its globular clusters (GCs). The dark matter content of Fornax is enough to induce strong dynamical friction in the motion of GCs and the timescale of orbital decay is shorter than the age of GCs. However, there are several GCs within Fornax that are far from its center.  
 
We can provide certain estimates for the influence of MW tides on the dynamical friction experienced by the GCs. To do so, it is first necessary to check the validity of our general approximation scheme. Our method works if the dynamical timescale within Fornax is much shorter than its orbital timescale about MW. Fornax's distance form MW is $\mathbb{R}=147\pm 12 \,$kpc, its total luminosity is $L= (1.4\pm 0.4)\times 10^7\, L_{\odot}$ and its half-light radius is $r_{\text{half}}=668\pm 34\,$pc \cite{Walker:2009zp}. On the other hand, the stellar mass-to-light ratio of Fornax is $\Upsilon_*\simeq 4.6\, \Upsilon_{\odot}$ \cite{Penarrubia:2007zx}. The baryonic matter density of Fornax is well fitted by a Plummer sphere~\cite{Walker:2009zp}, namely, 
 \begin{equation}
 \rho_b(r)=\frac{3 L \Upsilon_*}{4\pi r_{\text{half}}^3}\Big(1+\frac{r^2}{r_{\text{half}}^2}\Big)^{-\frac{5}{2}}\,.
 \label{G1}
 \end{equation}
 Finally, the dark matter content of Fornax can be described by  universal two-parameter mass profiles presented in~\cite{Walker:2009zp} for dSphs.  There are two different halo models for Fornax, a cuspy NFW halo and a cored one; for the present purpose, it does not matter which one is used. We employ the NFW dark matter halo:
 \begin{equation}
 \rho_d(r)=0.368\frac{\, V_{\text{max}}^2}{G\,r_h^2}\Big(\frac{r_h}{r}\Big)\Big(1+\frac{r}{r_h}\Big)^{-2}\,,
 \label{G2}
 \end{equation}
where $r_h$ is the halo lengthscale and $V_{\text{max}}$ is the maximum circular velocity associated with this halo. In the case of Fornax, these parameters are given by $V_{\text{max}}=18\pm 1\,\text{km s}^{-1}$ and $r_h = 795\,$pc~\cite{Walker:2009zp}. Both densities~\eqref{G1} and~\eqref{G2} can be integrated to find the corresponding masses, i.e., $M_b(r)$ and $M_d(r)$, enclosed within radius $r$. Let us define the baryonic mean density as $\langle\rho_b\rangle=3 M_b(r_{\text{half}})/(4\pi r_{\text{half}}^3)$. Using the above-mentioned quantities we find $\langle\rho_b\rangle\simeq 1.8\times 10^{-2}\,M_{\odot}\,\text{pc}^{-3}$ and we can hence calculate the internal Fornax dynamical timescale, namely, 
\begin{equation}
t_{\text{dyn}}=\frac{1}{\sqrt{G\,\langle\rho_b\rangle}}\simeq 1.14\times 10^{8}\, \text{yr}\,.
\label{G3}
\end{equation}
This is consistent with the estimate reported in~\cite{hammer2018} using a different method. On the other hand, the Keplerian period of Fornax in orbit about the MW is given by
\begin{equation}
T =  2\pi\sqrt{\frac{\mathbb{R}^3}{G\,\mathbb{M}}}\simeq 3.87\times 10^9\,\text{yr}\,,
\label{G4}
\end{equation}
where  we have used $\mathbb{M}\simeq 2\times10^{12}\,M_{\odot}$ for the mass of the MW. Consequently, we have $T/t_{\text{dyn}}\simeq 33.8$. Notice that by taking into account the dark matter density in Eq.~\eqref{G3}, the ratio $T/t_{\text{dyn}}$ gets even larger. There are other possible methods to estimate $T$, such as using the Fornax orbital parameters from Gaia DR2 data \cite{Gaia:2018kkg}. However,  all these estimates indicate that $T/t_{\text{dyn}} \gg 1$ and hence our approximation scheme is reasonably valid in this case. 
 
We are now in a position to estimate the tidal influence of the MW on the dynamical friction within Fornax. More specifically, we compute 
\begin{equation}
\Psi_x = -\frac{1}{2} \,\Psi_z = \frac{\pi\,\lambda}{M'}\frac{\mathbb{T}(b_{\rm min}, u_0|_{b_{\rm min}})}{\mathbb{S}(b_{\rm min}, u_0|_{b_{\rm min}})}\,,
\label{G5}
\end{equation}
on the basis of Eqs.~\eqref{F37} and~\eqref{Z13}.
To do so, we take the mean value of the King model core radii of Fornax GCs as representative of $b_{\min}\simeq 4.11\,$pc~\cite{Boldrini:2019yvk}. Similarly, we use the mean mass of the GCs  for $M'\simeq M\simeq  1.91\times10^5\,M_{\odot}$. Let us recall here that the tidal radius  $R_0$ is independent of the size of the binary system and is purely determined by the net two-body mass and the characteristics of the external source. For all interactions between a GC  and the stars of the background medium, we get essentially the same tidal radius $R_0$. That is, 
\begin{equation}
(u_0^2+b^2)^{1/2} < R_0 = \left(\frac{M'}{\mathbb{M}}\right)^{1/3}\,\mathbb{R}\,,
\label{G6}
\end{equation}
where for Fornax, $R_0 = 0.672$ kpc. On the other hand, the characteristic size of the Fornax galaxy is given by $D=2\, r_{200}$, where $r_{200}$ is the radius at which $\rho_b+\rho_d=200\, \rho_{\text{crit}}$. 
Here,  $\rho_{\text{crit}}=3H_0^2/(8\pi G)$ is the cosmic critical density and $H_0$ is the Hubble constant, which we assume is given by $H_0= 70 \, \text{km}\,\text{s}^{-1} \text{Mpc}^{-1}$. Using Eqs.~\eqref{G1} and~\eqref{G2}, one can numerically determine $r_{200}$; hence,  we find $D \approx 18\,$kpc.  Therefore, $D>R_0$ and we find from Eq.~\eqref{G5} that $u_0|_{b_{\rm min}}$ is at most  $\approx 0.672$ kpc. The determination of dynamical friction force in Fornax is not a simple task without implementing an appropriate distribution function. On the other hand, the ratios $\Psi_x$ and $\Psi_z$ are helpful in estimating the significance of tides in our problem. Given all the parameters discussed above, these ratios are $\Psi_x \approx 0.07$ and $\Psi_z \approx -0.14$. 
Therefore, for motion in the $(x,y)$ plane the tidal effects enhance dynamical friction, while for motion along the $z$ axis tidal stretching diminishes dynamical friction. For motion along an arbitrary direction with polar angle $\theta$, the corresponding ratio is given by $\Psi_z P_2(\cos\theta)$. Thus far, our preliminary analysis for Fornax implies that tidal effects would affect dynamical friction by a factor of around $10\,\%$; however, an appropriate distribution function is required to reach firm conclusions. Nonetheless, this conclusion is consistent with the observations where no sign of tidal disturbances is found in the stellar component within Fornax~\cite{Wang:2016qol, DES:2018jtu}. Furthermore, our result is consistent with the $N$-body simulations claiming that the internal kinematics of Fornax is only mildly influenced by the tidal effects of MW~\cite{BSN}.

\section{Discussion}

We have presented an approximation scheme based on expansion in powers of the gravitational constant $G$  to study the Chandrasekhar dynamical friction force in the presence of tides. The tidal interaction is considered within the distant tide linear perturbation approach. To extend our results to a system of stars, we need to determine the distribution function of the stars when tides are present. However, in this initial analytic study of dynamical friction in the presence of tides, we neglect the influence of tides on the phase space distribution of stars.  The self-gravity of the stellar system is neglected as well. Depending on circumstances, tidal forces can strengthen or weaken the dynamical friction force. In connection with possible astrophysical applications of our results, we consider the influence of Milky Way on Fornax dwarf galaxy. In this case, preliminary estimates suggest that the effect of Galactic tides on dynamical friction within Fornax dSph galaxy could be around $10\,\%$.

\section*{Acknowledgments}

M.R. is grateful to Elena Asencio and Indranil Banik for sharing their unpublished research on the tidal stability of Fornax cluster dwarf
galaxies. The work of M.R. has been supported by the Ferdowsi University of Mashhad.

\appendix

\section{Influence of Tides on a Gravitational $N$-Body System}

We consider a Newtonian astronomical system of $N$ bodies with inertial masses $M_{\alpha},~ \alpha = 1, 2, ..., N$, in a background Cartesian coordinate system $\mathbf{X} = (X, Y, Z)$ as in Figure~\ref{tide}.  The Newtonian gravitational force on $M_\alpha$ due to $M_\beta$ is
\begin{equation}\label{A1} 
\mathbf{F}_{\alpha \beta} = \frac{G\,M_\alpha\, M_\beta\,(\mathbf{X}_\beta - \mathbf{X}_\alpha)}{|\mathbf{X}_\beta - \mathbf{X}_\alpha|^3}\,. 
\end{equation} 
The $N$-body system is placed in the exterior gravitational field of an external source with potential $\Phi(\mathbf{X})$, so that $\nabla^2 \Phi = 0$ at the $N$-body system. The equation of motion for $M_\alpha$ is
\begin{equation}\label{A2} 
M_\alpha\,\frac{d^2X_{\alpha}^i}{dt^2} = - M_\alpha\, \delta^{ij}\nabla_j \Phi(\mathbf{X}_\alpha) +\sum_{\beta \ne \alpha}\,F^i_{\alpha \beta} \,. 
\end{equation}
We define the center of mass of the $N$-body system in the standard manner, namely, 
\begin{equation}\label{A3} 
\mathbf{X}_{CM} = \frac{\sum_{\alpha}\,M_\alpha\, \mathbf{X}_{\alpha}}{\sum_{\alpha} M_\alpha}\,.
\end{equation}
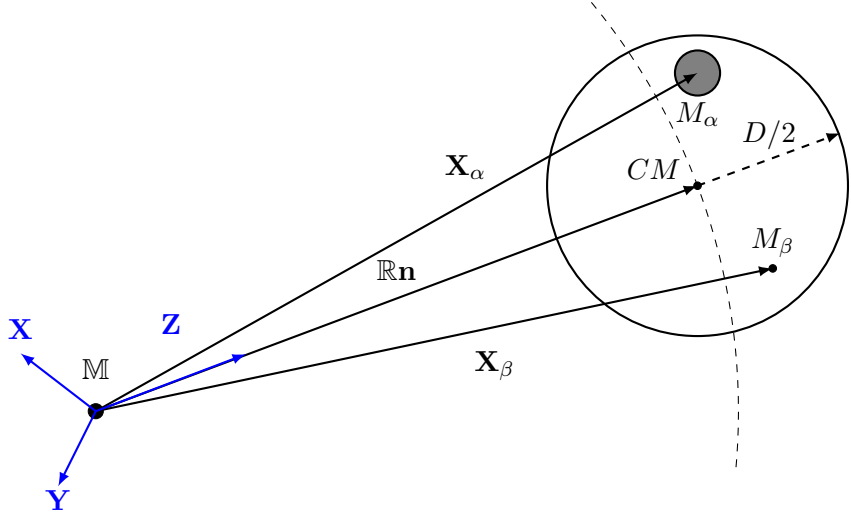
\begin{figure}
\centering
\begin{tikzpicture}
\draw[thick] (2,2) circle (2cm)node[left=0.58cm,above=-0.05cm]{$CM$};
\filldraw[fill=gray,thick] (2.,3.5) circle (0.3cm) node[above=-0.85cm]{$M_{\alpha}$};
\filldraw[fill=black] (2,2) circle (0.05cm);
\filldraw[fill=black] (3,0.9) circle (0.05cm)node[above=0.02cm]{$M_{\beta}$};
\filldraw[fill=black] (-6,-1) circle (0.1cm)node[above=0.3cm]{$\mathbb{M}$};
\draw[->, thick,  arrows={-latex}](-6,-1) -- (2.,3.5)node [midway,above=1cm,right=0.5cm] {$\mathbf{X}_{\alpha}$};
\draw[->, thick,  arrows={-latex}][thick](-6,-1) -- (3,0.9)node [midway,right=0.8cm,below=0.01cm] {$\mathbf{X}_{\beta}$};
\draw[->, thick,  arrows={-latex}][thick](-6,-1) -- (2,2)node [midway,above=0.1cm] {$\mathbb{R}\mathbf{n}$};
\draw[->, thick,  blue,arrows={-latex}][thick](-6,-1) -- (-4,-0.24)node [midway,above=0.5cm] {$\mathbf{Z}$};
\draw[->, thick,  blue,arrows={-latex}][thick](-6,-1) -- (-7,-0.23)node [above=0.05cm] {$\mathbf{X}$};
\draw[->, thick,  blue,arrows={-latex}][thick](-6,-1) -- (-6.5,-2)node [above=-0.45cm] {$\mathbf{Y}$};
\draw[->, thick, dashed,  arrows={-latex}][thick](2,2) -- (3.9,2.7)node [midway,above] {$D/2$};
    \draw[dashed] (-6,-1)+(-5:8.544cm) arc[start angle=-5, end angle=40, radius=8.544cm];
\end{tikzpicture}
\caption{Schematic illustration of the tidal interaction under consideration in this paper: A stellar system with characteristic size $D$ is tidally influenced by a distant mass $\mathbb{M}$.}\label{tide}
\end{figure}
Let us now introduce the approximation that the linear size of the $N$-body system $D$ is much smaller than $\mathbb{R}$, the distance to the source; that is $D\ll \mathbb{R}$. Therefore, we make a first-order tidal approximation
\begin{equation}\label{A4} 
\nabla_i \Phi(\mathbf{X}_\alpha) \approx \nabla_i \Phi(\mathbf{X}_{CM}) + \nabla_i\,\nabla_j\,\Phi(\mathbf{X}_{CM}) \,(X_\alpha^j - X_{CM}^j)\,. 
\end{equation}
It is useful to define the symmetric and traceless tidal matrix $K_{ij}$,
\begin{equation}\label{A5} 
K_{ij} = \frac{\partial^2 \Phi}{\partial X^i\,\partial X^j}\,, \qquad  K_{ij} = K_{ji}\,, \qquad \rm{tr}(K_{ij}) = 0\,.
\end{equation}
Moreover, we define
\begin{equation}\label{A6} 
x_{\alpha}^i = X_\alpha^i - X_{CM}^i\,,
\end{equation}
so that Eq.~\eqref{A2} now takes the form
\begin{equation}\label{A7} 
M_\alpha\,\frac{d^2x_{\alpha}^i}{dt^2} + M_\alpha\,\frac{d^2X_{CM}^i}{dt^2} = - M_\alpha\,\delta^{ij} \nabla_j \Phi(\mathbf{X}_{CM}) - M_\alpha\,K^{i}{}_{j}(\mathbf{X}_{CM})\,x_{\alpha}^j + \sum_{\beta \ne \alpha}\,F^i_{\alpha \beta} \,. 
\end{equation}
If we now sum this equation  over $\alpha$ and use $\sum_{\alpha}M_\alpha\,\mathbf{x}_{\alpha} = 0$, we find to first order in the tides
\begin{equation}\label{A8} 
\frac{d^2\mathbf{X}_{CM}}{dt^2} = - \nabla \Phi(\mathbf{X}_{CM})\,, 
\end{equation}
which describes the motion of the center of mass of the system about the source as though the tides never existed. This is, of course, true at the linear order in tidal perturbation.  We note that $\mathbf{X}_{CM}$ is a function of time $t$, since the center of mass of the system orbits about the distant mass $\mathbb{M}$. 
Combining Eqs.~\eqref{A7} and~\eqref{A8}, we find the important result that 
\begin{equation}\label{A9} 
M_\alpha\,\frac{d^2x_{\alpha}^i}{dt^2}  =  - M_\alpha\,K^{i}{}_{j}(\mathbf{X}_{CM})\,x_{\alpha}^j + \sum_{\beta \ne \alpha}\,F^i_{\alpha \beta}\,, 
\end{equation}
where in $F^i_{\alpha \beta}$, we have $\mathbf{X}_\beta - \mathbf{X}_\alpha = \mathbf{x}_\beta - \mathbf{x}_\alpha$. 

Let us now derive the energy equation for the internal motions of the $N$-body system. To this end, we introduce $\mathbf{v_{\alpha}} = d\mathbf{x_{\alpha}}/dt$ in Eq.~\eqref{A9} in the standard manner and obtain
\begin{equation}\label{A10} 
M_\alpha\,\mathbf{v}_\alpha \cdot \frac{d\mathbf{v}_\alpha}{dt}  =- M_\alpha\,K_{ij}(\mathbf{X}_{CM})\,v_{\alpha}^i\,x_{\alpha}^j +  \sum_{\beta \ne \alpha}\,\mathbf{v}_\alpha \cdot \mathbf{F}_{\alpha \beta}\,. 
\end{equation}
The total internal energy $\mathcal{E}$ is the sum of internal kinetic and potential energies,
\begin{equation}\label{A11} 
\mathcal{E} = \mathcal{K} + \mathcal{W}\,, \qquad  \mathcal{K} =\frac{1}{2} \sum_{\alpha}M_\alpha\,v_\alpha^2\,, \qquad \mathcal{W} = - \frac{1}{2}\, \sum_{\alpha,\beta}^{\prime}\,\frac{G\,M_\alpha\, M_\beta}{|\mathbf{x}_\alpha - \mathbf{x}_\beta|}\,, 
\end{equation}
where a prime over the summation sign indicates that $\alpha \ne \beta$ in the sum. Then, Eq.~\eqref{A10} implies
\begin{equation}\label{A12} 
 \frac{d\mathcal{E}}{dt}  = - K_{ij}(\mathbf{X}_{CM})\,\sum_{\alpha} M_\alpha\,v_{\alpha}^i\,x_{\alpha}^j = - \frac{1}{6} K_{ij} \frac{dQ^{ij}}{dt}\,,
\end{equation}
where $Q^{ij}$ is the symmetric and traceless internal quadrupole tensor of the system given by
\begin{equation}\label{A13} 
Q^{ij} = \sum_{\alpha} M_\alpha\,( 3 x_{\alpha}^i \,x_{\alpha}^j - x_{\alpha}^2 \,\delta^{ij})\,.
\end{equation}
Let us note that we can write
\begin{equation}\label{A14} 
 \frac{d}{dt}\left(\mathcal{E} + \frac{1}{6} K_{ij} \,Q^{ij}\right) =  \frac{1}{6} \frac{dK_{ij}}{dt}\,Q^{ij}\,.
\end{equation}
The temporal dependence of $K_{ij}$ has to do with the motion of the center of mass of the system, which is typically very slow compared to motions within the $N$-body system. During the fast internal motions of the system, the slow motion of the center of mass may be considered to be approximately uniform; that is, it may be a reasonable approximation in some cases to neglect the temporal dependence of the slow motion. Then,  Eq.~\eqref{A14} implies that we have a tidally disturbed $N$-body system in the first order of tidal perturbation that is  approximately conservative; 
that is, the sum total of kinetic plus potential plus tidal energies remains constant in time. 

Similarly, the internal angular momentum of the system can be defined via
\begin{equation}\label{A15} 
\mathcal{L}_i = \epsilon_{ijk} \sum_{\alpha}M_\alpha\,x_{\alpha}^j\,\frac{dx_{\alpha}^k}{dt}\, 
\end{equation}
and equation of motion~\eqref{A9} then implies
\begin{equation}\label{A16} 
\frac{d\mathcal{L}^i}{dt}  = \frac{1}{3} \epsilon^{ijk} K_{j}{}^{l} \,Q_{lk}\,.
\end{equation} 

For astrophysical applications, this approach can be extended to include the \emph{virial theorem}~\cite{BT} as follows. Let us define the quantities related to the moment of inertia of the system, namely, 
\begin{equation}\label{A17} 
\mathbb{I} = \frac{1}{2} \sum_{\alpha} M_\alpha\, x_{\alpha}^2\,, \quad  \frac{d\,\mathbb{I}}{dt} =  \sum_{\alpha} M_\alpha\, \mathbf{x}_{\alpha} \cdot \mathbf{v}_{\alpha}\,, \quad  \frac{d^2\mathbb{I}}{dt^2} = 2\mathcal{K} +\sum_{\alpha} M_\alpha\, \mathbf{x}_{\alpha} \cdot \frac{d\mathbf{v}_{\alpha}}{dt}\,.
\end{equation}
Then, from  Eq.~\eqref{A9} we get 
\begin{equation}\label{A18} 
M_\alpha\,\mathbf{x}_{\alpha} \cdot \frac{d\mathbf{v}_{\alpha}}{dt}  = - K_{ij}(\mathbf{X}_{CM}(t))\,M_\alpha\,x_{\alpha}^i\,x_{\alpha}^j +\sum_{\beta \ne \alpha}\,\frac{G\,M_\alpha\, M_\beta\,(\mathbf{x}_\beta - \mathbf{x}_\alpha)\cdot \mathbf{x}_{\alpha}}{|\mathbf{x}_\beta - \mathbf{x}_\alpha|^3}\,. 
\end{equation}
We sum this equation over $\alpha$ and write the sum again with $\alpha$ and $\beta$ exchanged. Adding the resulting equations we finally get in the standard manner
\begin{equation}\label{A19} 
\sum_{\alpha} M_\alpha\, \mathbf{x}_{\alpha} \cdot \frac{d\mathbf{v}_{\alpha}}{dt} = \mathcal{W} - \frac{1}{3}\,K_{ij}(\mathbf{X}_{CM}(t))\,Q^{ij}\,,
\end{equation}
which is equal to $\ddot{\mathbb{I}} - 2\mathcal{K}$ by Eq.~\eqref{A17}. Finally, assuming the ``fast" system relaxes over timescales short compared to the timescale of the ``slow" center-of-mass motion, we can average our result over time.  Assuming that the average of $\ddot{\mathbb{I}}$ over time vanishes~\cite{L+L}, we finally have the result
\begin{equation}\label{A20} 
 2\, < \mathcal{K}> + <\mathcal{W}> \, = \frac{1}{3}\,K_{ij}\,<Q^{ij}>\,.
\end{equation}

\section{Useful Integrals}

The integration of equations of motion in Section III is simplified using the indefinite integrals given below.  Let us assume $F = (x^2+b^2)^{1/2}$; then, 
\begin{equation}\label{B1} 
\int \frac{dx}{F}  = \ln(F + x)\,, \qquad \int \ln(F + x)\,dx = x\, \ln(F + x) - F\,,
\end{equation}
\begin{equation}\label{B2} 
\int  \frac{dx}{F^3}  = \frac{x}{b^2\,F}\,, \qquad  \int  \frac{dx}{F^5}  = \frac{1}{b^4F^3}\,x \left(\frac{2}{3} x^2 + b^2\right)\,,
\end{equation}
\begin{equation}\label{B3} 
 \int \frac{xdx}{F}  =  F\,, \qquad \int \frac{xdx}{F^3}  = - \frac{1}{F}\,, \qquad \int \frac{x^2dx}{F^5} = \frac{x^3}{3b^2F^3}\,,
\end{equation}
\begin{equation}\label{B4} 
 \int (\ln x) x dx  =  \frac{1}{2} x^2 \ln x - \frac{1}{4} x^2\,, 
\end{equation}
\begin{equation}\label{B5} 
\int \ln(F - x) b\, db = \frac{1}{2} b^2 \ln(F-x) -\frac{1}{2}x\,F -\frac{1}{4} b^2\,,
\end{equation}
\begin{equation}\label{B6} 
 \int \frac{(\ln b) b\, db}{F^3}  = - \frac{\ln b}{F} +\frac{1}{2x}\, \ln\left(\frac{F-x}{F+x}\right)\,,
\end{equation}
\begin{equation}\label{B7} 
 \int \frac{\ln (F + x)}{F^3} b\, db = -\frac{\ln (F + x)}{F} +\frac{1}{x}[\ln F - \ln (F+x)]\,.
\end{equation}

We also need to evaluate the definite integral
\begin{equation}\label{B8} 
I = \int_{-x_0}^{x_0} \frac{\ln(F + x)}{F^3}\,dx  = \int_{-x_0}^{x_0} \frac{\ln(F - x)}{F^3}\,dx\,. 
\end{equation}
From $\ln(F + x) + \ln(F - x) = 2\ln b$, we conclude
\begin{equation}\label{B9} 
I = \ln b \,\int_{-x_0}^{x_0} \frac{dx}{F^3}  = 2 \frac{\ln b}{b^2} \frac{x_0}{F_0}\,,
\end{equation}
where $F_0 = F(x_0, b)$.  Similarly,
\begin{equation}\label{B10} 
\int_{-x_0}^{x_0} \frac{\ln(F + x)}{F^5}\,dx  = 2 \frac{\ln b}{b^4} \frac{x_0}{F_0^3}\, \left(\frac{2}{3} x_0^2 + b^2\right)\,.
\end{equation}
The corresponding definite integrals can be easily evaluated via integration by parts using Eq.~\eqref{B2}.

\section{Dynamical Friction and Tidal Interactions in Nonlocal Gravity}

In a previous paper~\cite{Roshan:2021ljs}, we extended Chandrasekhar's formula for dynamical friction to the Newtonian regime of nonlocal gravity theory and briefly studied its implications for barred spiral galaxies. Nonlocal gravity (NLG) is a classical nonlocal generalization of Einstein's theory of gravitation that takes the past history of the gravitational field into account. The gravitational field in NLG is local, but satisfies partial integro-differential field equations. Moreover, NLG has been constructed in close analogy with the nonlocal electrodynamics of media. It turns out that such a classical nonlocal aspect of the gravitational interaction simulates dark matter. A detailed description of NLG is contained in~\cite{BMB}. It seems worthwhile to indicate briefly how the formal results of the present work can carry over to NLG. To this end, we must first describe the analogue of the Newtonian inverse-square law in NLG. 

In the Newtonian limit, NLG in effect involves the standard nonrelativistic gravitational force
\begin{equation}\label{C1}
\mathbf{F}_{\rm NLG}(\mathbf{x}) = - m \nabla\Phi_{\rm NLG}(\mathbf{x})\,
\end{equation}
on a test particle of inertial mass $m$ in the gravitational potential $\Phi_{\rm NLG}(\mathbf{x})$, which satisfies the nonlocal Poisson equation that can be expressed in the form
\begin{equation}\label{C2} 
\nabla^2\Phi_{\rm NLG} = 4\pi G\,(\rho+\rho_D)\,, \qquad \rho_D(\mathbf{x})=\int q(\mathbf{x}-\mathbf{y}) \rho(\mathbf{y})\,d^3y\,.
\end{equation}
Here, $\rho$ is the density of matter and $\rho_D$ is the density of effective dark matter in NLG. In this theory, what appears as dark matter in astrophysics and cosmology is in reality the nonlocal aspect of gravity itself and its density is given by the convolution of a certain reciprocal kernel $q$ with the density of matter $\rho$. The field equations of nonlocal gravity in the Newtonian regime of the theory reduce to the nonlocal Poisson equation~\eqref{C2} provided the functions involved be smooth and satisfy certain reasonable mathematical properties~\cite{BMB}. Moreover, the reciprocal kernel $q$ must be determined on the basis of observational data. 

To simplify matters, let us assume that $q(\mathbf{x} - \mathbf{y})$ is spherically symmetric; then, for $q(r)$, $r = |\mathbf{x} - \mathbf{y}|$, two possible forms have been discussed in detail. These are~\cite{BMB}   
\begin{equation}\label{C3}
 q_1 (r) = \frac{1}{4\pi \lambda_0} \,\frac{1+\mu_0\, (a_0+r)}{r\,(a_0 + r)}\,e^{-\mu_0\,r}\,, \quad q_2 (r) = \frac{r}{a_0 + r}\,q_1(r)\,,
\end{equation}  
which contain three constant parameters. The basic NLG lengthscale is determined by $\lambda_0$, since nonlocality disappears when $\lambda_0$ tends to infinity; moreover, $a_0$ moderates the short distance behavior of the kernel and $\mu_0$ is the ``Yukawa" parameter.  It proves useful to define a dimensionless parameter $\alpha_0 = 2/(\lambda_0\,\mu_0)$. Solar system data provide a lower bound for $a_0$, namely, $a_0 > 10^{14}$~cm~\cite{Chicone:2015coa}. With $a_0 = 0$, $q_1 = q_2$ and the rotation curves of nearby spiral galaxies can be used to find $\mu_0$ and $\lambda_0$. Indeed, observational data regarding nearby spiral galaxies and clusters of galaxies are consistent with~\cite{Rahvar:2014yta} 
\begin{equation}\label{C4}
\alpha_0 = 10.94 \pm 2.56\,,\quad \mu_0 = 0.059 \pm 0.028~{\rm kpc}^{-1}\,, \quad  \lambda_0 = \frac{2}{\alpha_0\,\mu_0} \approx 3\pm 2~{\rm kpc}\,.
\end{equation} 

It is straightforward to calculate the force of gravity according to NLG on a point mass $m$ due to another point mass $m'$ at position $\mathbf{r}$. The result is a modification of Newton's law of universal  gravitational attraction given by
\begin{equation}\label{C5}
 \mathbf{F}_{\rm NLG} =  \frac{Gmm' \mathbf{r}}{r^3} [1+ \Delta(r)]\,.
\end{equation}
A detailed physical interpretation of this result is contained in~\cite{Roshan:2021ljs}. In Eq.~\eqref{C5},  the net contribution of the effective dark matter is contained in $\Delta(r) \ge 0$. The function $\Delta(r)$ starts from zero at $r = 0$, increases monotonically with increasing $r$ and approaches $\alpha_0 \,w$ asymptotically as $r \to \infty$.  Here, $w$ is a positive constant such that $w = 1$ for $a_0 = 0$ and for $a_0 > 0$, $w$ depends upon whether the reciprocal kernel is chosen to be $q_1$ or $q_2$; however, for reasonable values of $a_0$ less than a few parsecs, $w$ is very close to unity. Henceforth, we ignore the deviation of $w$ from unity. That is, NLG in the Newtonian regime is such that the magnitude of force of gravity asymptotically approaches Newtonian gravity  except that now the effective constant of gravitation is about an order of magnitude larger than the standard Newtonian constant of gravitation $G$. 

With these introductory remarks about $ \mathbf{F}_{\rm NLG}$, we are now in a position to indicate how our Newtonian approach in this paper is affected by the presence of $\Delta(r)$. In Eqs.~\eqref{T1}--\eqref{T2} and~\eqref{T8}--\eqref{T10}, the internal forces get multiplied by $[1+\Delta(r)]$, while the external forces get multiplied by $[1+\Delta(\mathbb{R})]$. In the distant tide approximation, if $\mathbb{R} \gg \mu_0^{-1}$, then $\Delta(\mathbb{R}) \approx \alpha_0$. In this case, the tidal radius is smaller than the Newtonian $R_0$ given by Eq.~\eqref{T13}. Extending these considerations to dynamical friction, it has been shown in~\cite{Roshan:2021ljs} that the nontidal ``Chandrasekhar" dynamical friction force gets multiplied by $(1+\hat{\eta})^2$, where $\hat{\eta}$, $0 < \hat{\eta} <  \alpha_0$, is a constant that depends on the parameters of the stellar system and has been discussed in detail in Section IV-A of~\cite{Roshan:2021ljs}. In the present case, we find that in Eq.~\eqref{W16} the nontidal dynamical friction force gets multiplied by $(1+\hat{\eta})^2$ as before, while the tidal part of dynamical friction force gets multiplied by  $(1+\hat{\eta})(1+\alpha_0)$. Finally, the matter density in Eq.~\eqref{W16} for dynamical friction would then simply refer to the baryonic density in this case, as there is no actual dark matter in NLG.

\end{document}